\documentclass[runningheads]{llncs}

\usepackage[T1]{fontenc}
\usepackage{graphicx}
\usepackage{booktabs}
\usepackage{amsmath}
\usepackage{xcolor}
\usepackage{multirow}
\usepackage{diagbox}
\usepackage{dirtytalk}
\usepackage{academicons}
\usepackage{colortbl}
\usepackage[most]{tcolorbox}
\usepackage{enumitem}
\usepackage{xurl}
\usepackage{cite}
\usepackage{csquotes}
\usepackage{listings}
\usepackage[title]{appendix}
\usepackage{tablefootnote}
\usepackage{algorithm}
\usepackage{algpseudocode}
\usepackage{epigraph}
\usepackage{booktabs}
\usepackage{subfig}
\usepackage{shapepar}
\usepackage{float}
\usepackage[depth=3]{bookmark}
\usepackage{minitoc}
\usepackage{pifont}
\usepackage{orcidlink}
\usepackage{bm}

\definecolor{adversarial}{HTML}{971a1e}
\definecolor{color_a}{HTML}{893244}
\definecolor{color_b}{HTML}{4F404C}
\definecolor{color_c}{HTML}{007F7F}
\definecolor{color_d}{HTML}{6082B6}

\title{
    StegoStylo: Squelching Stylometric Scrutiny through Steganographic Stitching
}

\author{
    Robert Dilworth \orcidlink{0009-0005-5497-9810}
}

\authorrunning{
    Robert Dilworth
}
\titlerunning{
    StegoStylo
}

\institute{
    Department of Computer Science and Engineering, Mississippi State University, Mississippi State, Mississippi, USA\\
    \email{rkd103@msstate.edu}
}

\hypersetup{
    pdftitle={StegoStylo: Squelching Stylometric Scrutiny through Steganographic Stitching},
    pdfsubject={cs.CR, cs.CL, cs.IR},
    pdfauthor={Robert Dilworth},
    pdfkeywords={Adversarial Stylometry, Privacy, Unicode Steganography with Zero-Width Characters},
    colorlinks=true,
    linkcolor=color_a,
    citecolor=color_c,    
    urlcolor=color_d,
}

\begin{document}

\maketitle

\begin{abstract}

    Stylometry---the identification of an author through analysis of a text's style (i.e., authorship attribution)---serves many constructive purposes: it supports copyright and plagiarism investigations, aids detection of harmful content, offers exploratory cues for certain medical conditions (e.g., early signs of dementia or depression), provides historical context for literary works, and helps uncover misinformation and disinformation\footnote{The abstract mentions several benign use cases of stylometry. See the work of the following researchers as they relate to the positive implementation of stylometric analysis: plagiarism (academic-integrity investigations) - \textit{Albaqami et al.} \cite{Albaqami2026}; harmful-content detection (filtering phishing emails) - \textit{Opara et al.} \cite{Opara2025}; medical-condition discovery (predicting ADHD, depression, schizophrenia) - \textit{Barrios et al.} \cite{Barrios2023}, \textit{P\'erez et al.} \cite{Perez2022}, \textit{Rezaii et al.} \cite{Rezaii2019}; providing historical context (composer identification through musical stylometry) - \textit{Abrosimov et al.} \cite{Abrosimov2025}; and misinformation/disinformation (detecting fake news) - \textit{Yang et al.} \cite{Yang2024}.}. In contrast, when stylometry is employed as a tool for authorship verification---confirming whether a text truly originates from a claimed author---it can also be weaponized for malicious purposes. Techniques such as de-anonymization, re-identification, tracking, profiling, and downstream effects like censorship illustrate the privacy threats that stylometric analysis can enable. Building on these concerns, this paper further explores how adversarial stylometry combined with steganography can counteract stylometric analysis. We first present enhancements to our adversarial attack, \textsc{TraceTarnish}, providing stronger evidence of its capacity to confound stylometric systems and reduce their attribution and verification accuracy. Next, we examine how steganographic embedding can be fine-tuned to mask an author's stylistic fingerprint, quantifying the level of authorship obfuscation achievable as a function of the proportion of words altered with zero-width Unicode characters. Based on our findings, steganographic coverage of 33\% or higher seemingly ensures authorship obfuscation. Finally, we reflect on the ways stylometry can be used to undermine privacy and argue for the necessity of defensive tools like \textsc{TraceTarnish}.
    
    \keywords{
        Adversarial Stylometry \and Privacy \and Unicode Steganography with Zero-Width Characters
        }

\end{abstract}

\section{New Research Angles}
\label{sec:New_Research_Angles}

    \epigraph{\textcolor{adversarial}{Words can be like X-rays, if you use them properly---they'll go through anything. You read and you're pierced.}}{\textit{Brave New World \\ Aldous Huxley}}

    A key unanswered question from our preceding studies (\textit{Dilworth} \cite{AdversarialStylometry2025,TraceTarnish2025}) concerns the trade-off between anonymization quality and detectability. Specifically, how much must \textsc{TraceTarnish}---an adversarial stylometry attack that anonymizes text authorship---degrade a message to guarantee reliable anonymization? If perfect anonymization renders the text unreadable or nonsensical, its practical usefulness collapses regardless of resistance to detection.  

    To address this, we must shift focus toward quantifying the optimal amount of \say{poisoning.} A promising approach is to measure the computed stylistic score of an original text alongside various steganographic embedding transformations, thereby identifying the degree of alteration that maximally obscures authorship while preserving readability.

    Inspired by \textit{Poisonify} (\textit{Jordan} \cite{Jordan2025})---an adversarial noise attack that sabotages AI music generation---and \textit{HarmonyCloak} (\textit{Meerza et al.} \cite{Meerza2025})---a defensive technique that shields music from generative models---we recognize untapped potential in how stego text (the result of embedding a hidden message within cover text) interacts with stylometric systems.

    In previous work, we sought to disrupt text-based pattern matching. For instance, if you wanted to search for a specific word within all the documents on your device, you could use the command-line utility \texttt{grep}\footnote{\texttt{grep}'s manual page (\href{https://www.man7.org/linux/man-pages/man1/grep.1.html}{man page}) describes its functionality as a utility that \say{prints lines that match patterns.}} to scour your files for a particular string or regular expression. All that is needed to thwart this search is to ensure that the utility never finds the pattern. One way to accomplish that is to \textit{inject} a zero-width Unicode character as an infix\footnote{An infix, as defined by the \href{https://www.merriam-webster.com/dictionary/infix}{Merriam-Webster} dictionary, is \say{a[n]\dots affix appearing in the body of a word.}} within the target string contained in a larger document. This is satisfactory; \texttt{grep} fails to detect the pattern, but a question arises: what is the extent of this embedding? As a disclaimer---and given the subject matter---it would be irresponsible not to provide a means of detecting the existence of invisible Unicode characters within the \texttt{grep} framework, as previously proposed. For those interested---or concerned---see the following resource \cite{LinuxBash2025} for guidance on how to discover such characters.

    For context, the stylometric application of \texttt{grep} involves using the tool to analyze textual patterns---such as word frequency, punctuation usage, and sentence structure---to assess an author's unique writing style. Precomputed word lists, like the most common function words, are often used in stylometric analysis to identify and quantify specific vocabulary choices, allowing an investigator to detect stylistic similarities and differences between texts or authors.

    From a privacy standpoint, the intuition is that corrupting a word in a text that also appears in a word list---so that a search of the list for the corrupted word fails---undermines the \texttt{grep}-based stylometric scheme described earlier (and similar approaches). A question that emerges from this line of thinking is whether the hidden text can inflict damage on a stylometric system beyond merely causing pattern-recognition misses. Despite the tantalizing prospect, determining upper and lower bounds on the degree of encoding takes precedence; establishing these limits is essential before considering broader implications.

    The significance of this question is made more acute given the ramifications of stylometry. George Mikros, in their work (\textit{Mikros} \cite{Mikros2025}), emphasizes that stylometry has valuable because \say{language provides insights about its creator,} and \say{every individual possesses a fairly consistent and unique linguistic \say{fingerprint} (idolect) that can be observed in both spoken and written expressions.}

    As in our earlier work, our research aims to facilitate text-based interactions while disclosing as little information as possible about the participants---whether one-to-one (direct messages), one-to-many (publications, social media posts), etc.---and rendering each linguistic fingerprint less recognizable. This could, hypothetically, be achieved by processing all outgoing and incoming text with a tool like \textsc{TraceTarnish}, corrupting each participant's stylistic fingerprint.

    In lieu of a conventional breakdown of our paper's structure, we present a visual outline that better resonates with the essence of our content. As we delve into our work, the section headings we choose may divulge aspects of our identity best kept anonymous. Thus, we embrace the perfect opportunity to employ \textsc{TraceTarnish}. To illuminate the intricacies of this attack, a gentle suspension of disbelief may be required.

\begin{figure}[H]
        \centering
        \includegraphics[width=0.5\linewidth]{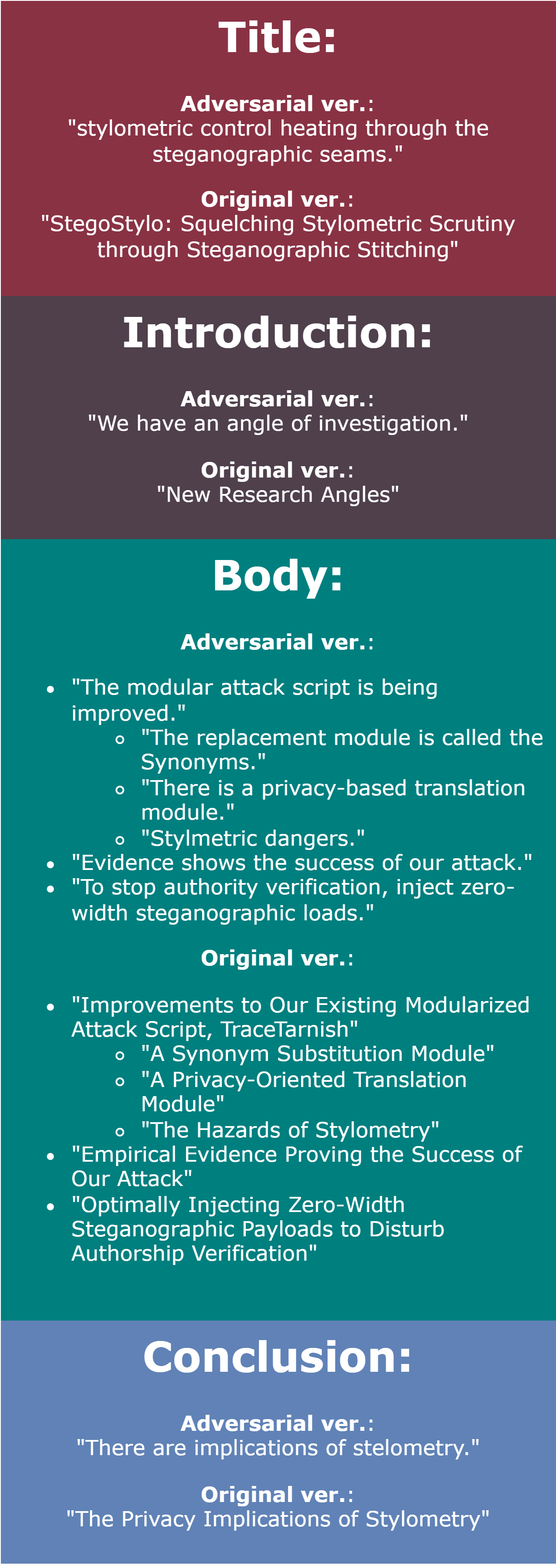}
        \caption{The structure of our paper, wherein the text has been compromised through the application of \textsc{TraceTarnish}.}
        \label{fig:paper_overview_corrupted}
    \end{figure}

    Might every facet of this paper, save for the figure's material, weave together a stylistic profile of its authors? The answer is a resounding yes. Yet, can you be certain that the text has not been poisoned with steganographic payloads meant to obscure such profiling? The answer rests in the shadows of debate. Regardless, we shall seize the opportunity nonetheless. See (\textbf{Figure \ref{fig:paper_overview_corrupted}}) for an overview of our paper's structure, where we showcase the various sections. However, the text has been corrupted using \textsc{TraceTarnish}.

\section{Improvements to Our Existing Modularized Attack Script, \textsc{TraceTarnish}}
\label{sec:Improvements_On_Script}

    While we are relatively content with the state of our code, we stumbled across repositories and articles that pertain to---or could be made relevant to---our project. Here, we spotlight a few of our findings that could either be incorporated into our project to improve it or writings that demonstrate the exigent perils of stylometry abuse.

    \subsection{A Synonym Substitution Module}
    \label{subsec:Synonym_Substitution_Module}

        In order to evaluate the integration of the code presented in \cite{LLMask2024}, we propose incorporating it as a distinct module within our attack script, positioned between the existing implementations of \say{adversarial translation} and \say{adversarial obfuscation.} This addition would involve rewriting with a focus on synonym substitution, while the existing module rewrites with an emphasis on paraphrasing; both are valid forms of \say{adversarial obfuscation.} 

        Synonym substitution, as used here, is the process of replacing a word in a text with another word that has the same or a very similar meaning, often to vary language, avoid repetition, or alter stylistic tone.

    \subsection{A Privacy-Oriented Translation Module}
    \label{subsec:Privacy_Translation_Module}

        While reviewing related publications, we discovered a privacy-focused machine translator \cite{translateLocally2021} designed to operate without an internet connection. The application can perform multiple consecutive round-trip translations of a source text to degrade its quality, thereby reducing the author's stylistic signature, all while maintaining privacy: the data never leaves the machine. The best way to protect data is to never release it (or to never store, generate, or transport it in the first place, but we digress), and the translator's development team acknowledges this.

    \subsection{The Hazards of Stylometry}
    \label{subsec:Hazards_Stylometry}

        To illustrate the immediate dangers of stylometry, we recommend reading Robert Lerner's excellent and illustrative blog post (\textit{Lerner} \cite{Lerner2025}). The central thesis is that there is no shortage of \say{tells}---think revealing gestures or expressions, irrespective of whether they occur online or in real life---that leave a trail. In isolation, they are innocuous; when combined and systematically scrutinized, the \say{crumbs of individuality} add up, for better or for worse.

\section{Empirical Evidence Proving the Success of Our Attack}
\label{sec:Empirical_Evidence}

    Before we go further, it is imperative that we first prove, beyond a reasonable doubt, that \textsc{TraceTarnish} does what it sets out to do. With that foundation laid, a typical deployment scenario for the attack---its ideal use case---unfolds as follows.
    
    Assume that there is a prolific whistleblower---someone who reports wrongdoing, such as immoral, unsafe, or unethical conduct within an organization, to promote accountability and transparency. The whistleblower has authored many publicly accessible works in an activist capacity. As a consequence of their activities and the notoriety they have garnered, a veritable spotlight, akin to a bounty, has been placed on their head. Any word about their activity, even something as mundane as a public sighting, throws the public---and, most importantly, their retaliators---into a frenzy. Given the numerous blows to public perception, reputational damage, and monetary loss, the whistleblower has engendered animus and ill-will from their targets, who now seek retribution. Dead or alive, the whistleblower's victims care not---whatever it takes, the whistleblower must be silenced.

    The whistleblower, with their newfound popularity (or perhaps infamy), begins taking steps to reduce their visibility. This drive to hide is spurred on by a newly unlocked fear that open-source intelligence (OSINT) and other adjacent disciplines could be used to identify, track, or expose them. An amateur had every right to feel dread and foreboding, but the whistleblower, hoping for the best but preparing for the worst, made prior arrangements to alleviate the cognitive burden of (essentially) being hunted. 
    
    What matters most is their path ahead. Should they persist in courting danger, they will need to obfuscate their communications, at least on a personal level. A database of their publications already exists; from this database, any future publications---whether a recount of their latest escapade or a simple chat with loved ones---could be tied back to them. The former is a foregone conclusion, as dire and uncontrollable circumstances dictate that writing of that ilk be clear and comprehensible. There is room for error, however, with the latter, which more often than not leaks the most identifiable information.

    The premise is that the whistleblower requires any future writings \textit{not} to be traceable or attributable to them. Considerations such as encryption are excluded from this scenario since someone---or something---somewhere will inevitably view the whistleblower's writing.

    In light of this, we devise a simple experiment. The whistleblower needs a tool that will ensure they are statistically unlikely to be identified as the author of any written communication, which we assume is penned and transmitted digitally. One such tool under consideration is \textsc{TraceTarnish}. To test whether it meets the whistleblower's specifications, we collect their writing, convert it to text files, and create a corpus for authorship verification. From this corpus, we take a random sample, exclude it from the set, and feed the sample to \textsc{TraceTarnish} to produce an adversarial output. The output is adversarial in that its properties should, in theory, thwart stylometric evaluation.

    At this stage, we have a corpus comprised of the whistleblower's works, the clean sample isolated from the corpus, and the adversarial transformation of the clean sample. With these pieces, we enlist the aid of the R package, \texttt{stylo} (\textit{Eder} \cite{Eder2018,Imposters2025}). In this scenario, we are most concerned with the authorship-verification features of the \texttt{imposters()} function, which takes a set of candidate author/text pairs and uses feature frequencies of words to calculate and assign a score between 0 and 1. Depending on the construction of three sets---(1) the text to be tested, (2) the collection of texts assumed to be written by the candidate author, and (3) a reference set of \say{imposters} that excludes the candidate works---the function's output can be used to ascertain authorship. The results can be interpreted as follows:

    \begin{itemize}
        \item[\ding{118}] \textbf{Score \( \boldsymbol{>} \) 0.5}: verification succeeded.
        \item[\ding{118}] \textbf{Score \( \boldsymbol{\approx} \) 0.39\( \boldsymbol{-} \)0.63}: suspicious; classifier was likely uncertain.
        \item[\ding{118}] \textbf{Score \( \boldsymbol{<} \) 0.5 (outside the suspicious range)}: verification failed.
    \end{itemize}

    To clarify how the \texttt{imposters()} function operates, the method proceeds in two stages. First, a first-order similarity score is computed between a set of texts \( D \) (known works of the candidate author) and a set of imposter texts \( P \) (works by other authors). Imposter texts are selected to match the query's genre, period, topic, and target audience, ensuring that the similarity comparison is meaningful. Using a normalized feature set---typically word-frequency or character n-gram counts---and a similarity metric such as Burrows's Delta, each text in \( D \) (more specifically, any pair of texts drawn from \( D \)) is compared to all texts in \( P \), yielding first-order scores.

    The second-order similarity step verifies these results. For every pair of texts \( (d_{i},d_{j}) \) from \( D \), the system computes their similiarity to a common query imposter \( p_{u} \) in \( P \); a second-order score is then recorded, indicating whether the similarity between \( d_{i} \) and \( d_{j} \) exceeds their respective similarites to any imposter in \( P \).

    Multiple second-order scores are aggregated by obtaining a median and compared to a predefined threshold \( (\theta) \). If the aggregated value exceeds the threshold, the system attributes the query to the candidate author; otherwise, it labels the authorship as unknown. The complete algorithm, as presented by Koppel and Seidman, is described in their 2017 paper (\textit{Koppel et al.} \cite{Koppel2017}), devoid of any errors that may have arisen from our oversimplification.

\begin{figure}[H]
        \centering
        \includegraphics[width=1\linewidth]{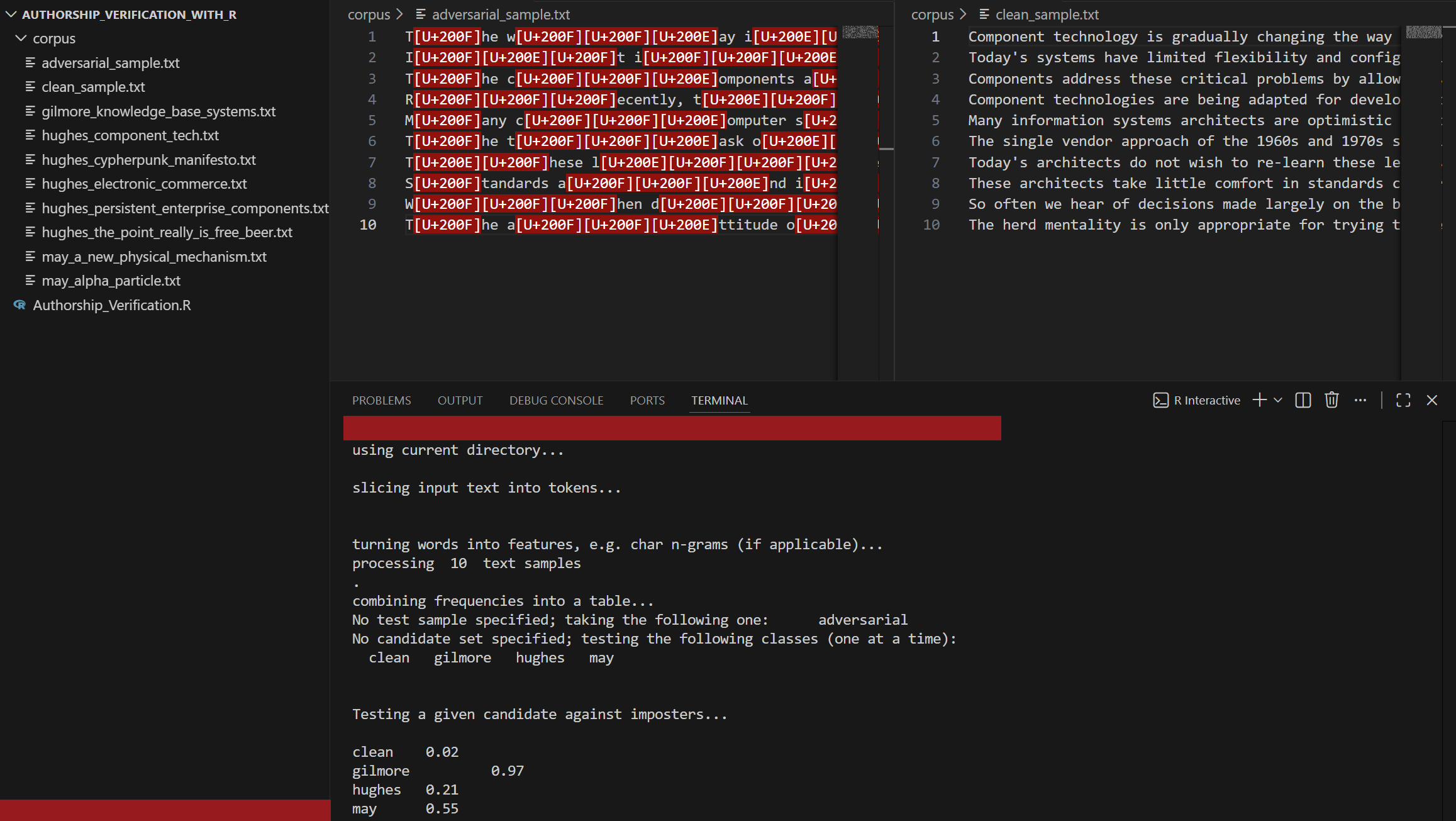}
        \caption{A screenshot of our IDE showing the scenario's corpus alongside the adversarial and clean samples processed with the \texttt{stylo} package's \texttt{imposters()} function; the resulting authorship-verification scores lend credence to our attack, demonstrating its validity within the adversarial stylometry domain.}
        \label{fig:stylo_imposter_results_adversarial}
    \end{figure}

    To break the immersion of the scenario, we repurposed the corpus from a previous study (see \cite{AdversarialStylometry2025}). Continuing the fourth-wall break, (\textbf{Figure \ref{fig:stylo_imposter_results_adversarial}}) showcases our corpus in the left-hand panel. The central, top-most panels contain the files of interest: \say{adversarial\_sample.txt} (left), which is the processed output of \say{clean\_sample.txt} (right). The bottom-most panel depicts the \texttt{imposters()} authorship-verification results. Note: the author of both samples, adversarial and clean, was Eric Hughes (referenced as \say{hughes} in the figure). Based on the results, John Gilmore is the most likely author of the adversarial test sample, which we know to be false. The second-most likely candidate is Timothy C. May; this determination is also false. Most importantly, Eric Hughes is indicated as the least likely author of the adversarial sample, which constitutes a success in this context. Hughes was the actual author of the adversarial text, yet he was not attributed as its author.

\begin{figure}[H]
        \centering
        \includegraphics[width=1\linewidth]{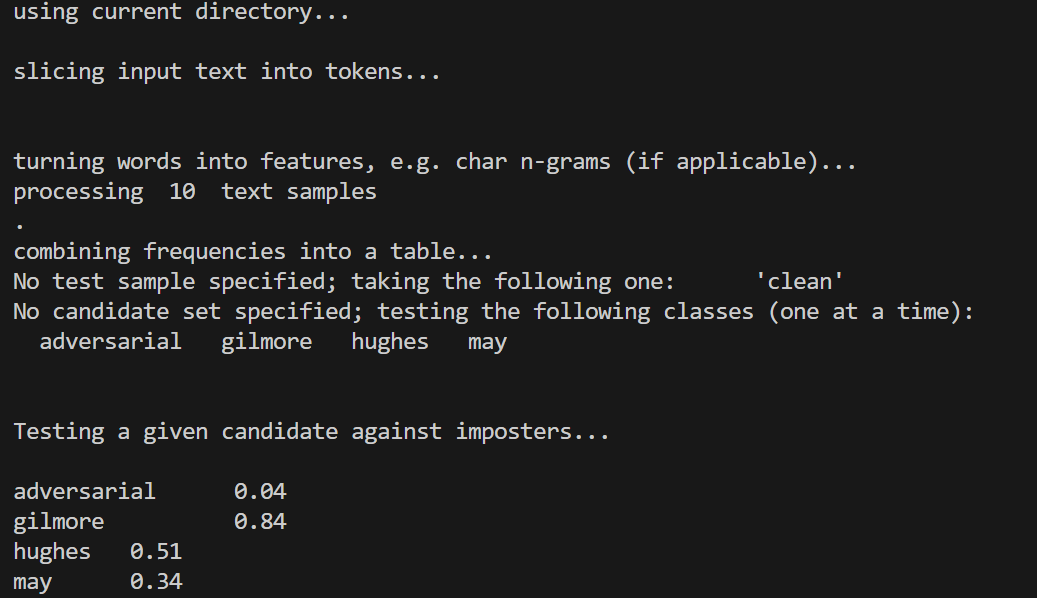}
        \caption{The output of the \texttt{stylo} package's \texttt{imposters()} function showing the inverse setting---here the clean sample is used as the test text, whereas in the previous figure the adversarial sample served that role.}
        \label{fig:stylo_imposter_results_clean}
    \end{figure}

    (\textbf{Figure \ref{fig:stylo_imposter_results_clean}}) further solidifies the result but also complicates matters slightly. Again, Gilmore is the strongest contender for the author, followed by Hughes. This is problematic, as we are (relatively) confident that Hughes wrote the sample, but this does not conflict with our interpretation criteria. Technically, both Gilmore's and Hughes's scores exceed the threshold for success, so Hughes remains suspect, although we would expect him to be the leading candidate.

    All things considered, and returning to the whistleblower example, a trial run of the tool led them to deem it suitable for their needs. While other aspects of their privacy remain jeopardized, they have taken appropriate measures to fool stylometric analyses. Given the severity of their situation, even a single mistake could spell their doom.

\section{Optimally Injecting Zero-Width Steganographic Payloads to Disturb Authorship Verification}
\label{sec:Optimally_Injecting}

    \epigraph{\textcolor{adversarial}{The\dots body persists although the component cells may change.}}{\textit{Brave New World \\ Aldous Huxley}}

    Within the context of our attack, we implement four adversarial stylometry techniques: Imitation, Injection, Translation, and Obfuscation. Based on our implementation, three of the four methods are all-or-nothing. Imitation: either you establish a fake author's profile and rewrite the text from that perspective, or you keep the text unchanged. Translation: either you translate the text from a source language to a target language and then back, or you do nothing. Obfuscation: either you paraphrase the text in its entirety, or you make no changes. However, Injection is a spectrum, which is precisely what we address in this section.

    Injection can be made incrementally. In parallel with the previous explanations, the question becomes: given an arbitrary set of words in a source text, to what degree will you insert steganographic payloads within elements of that set? Will it be 10\%? 50\%? 100\%? In terms of deceiving authorship verification, what is the optimal percentage of stego text to non-stego text? To answer this question, we replicated the setup of the previous whistleblower scenario with slight modifications.

\begin{figure}[H]
        \centering
        \includegraphics[width=1\linewidth]{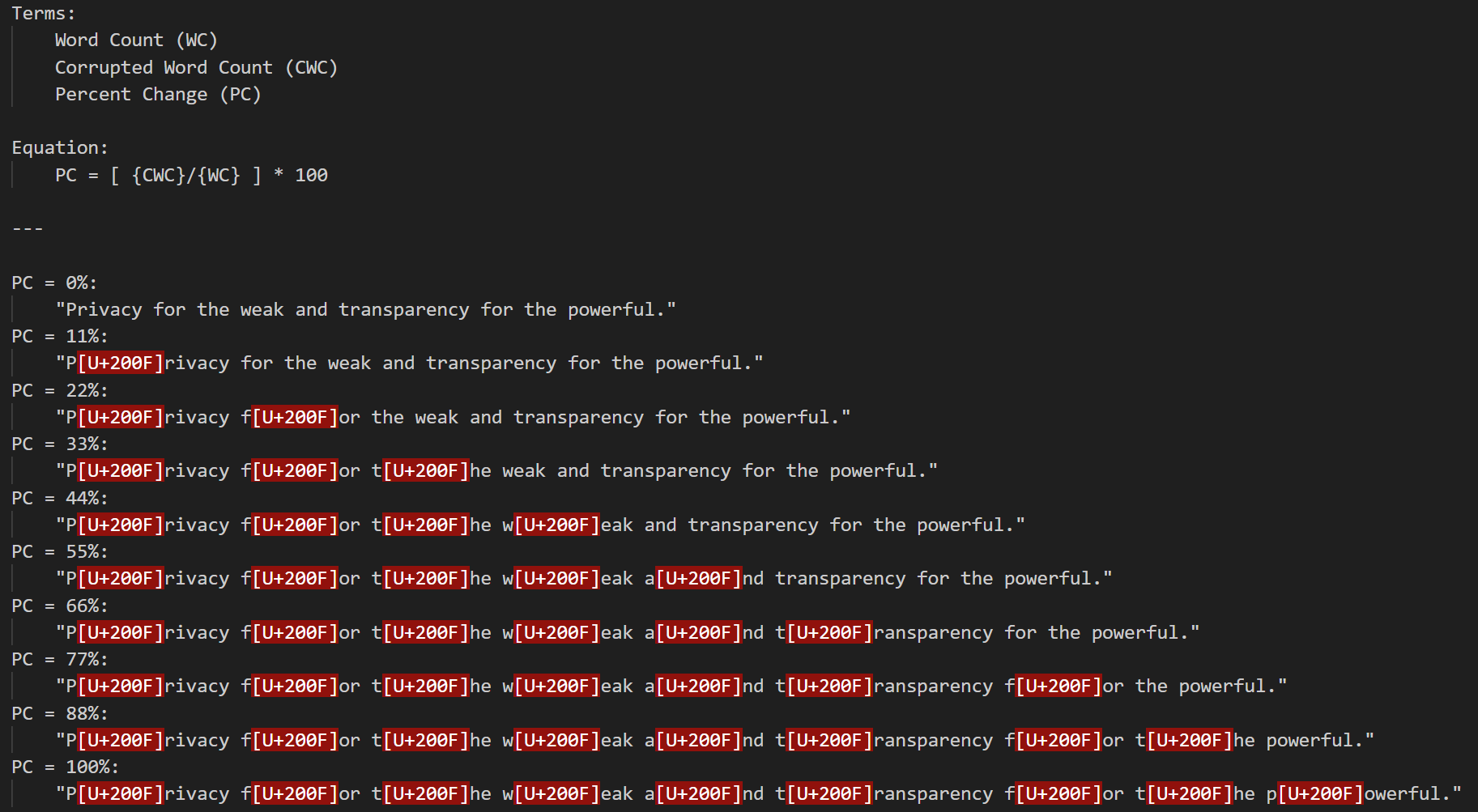}
        \caption{The sample texts comprising our new corpus assess injection optimality, addressing the questions: at what point does injection become effective, and what is the minimal degree of injection needed to achieve authorship obfuscation?}
        \label{fig:optimality_experiment_samples}
    \end{figure}

    For starters, we will use the same initial sample text, \say{Privacy for the weak and transparency for the powerful,} which contains nine words in total. This unaltered sample will serve as our control, representing 0\% change. For each of the nine words, we will iteratively inject a zero-width Unicode character, creating a new sample each time. The naming schema of the newly generated files will indicate the percentage of injection that has taken place, up to full injection (100\%). See (\textbf{Figure \ref{fig:optimality_experiment_samples}}) for the samples under consideration.

\begin{figure}[H]
        \centering
        \includegraphics[width=1\linewidth]{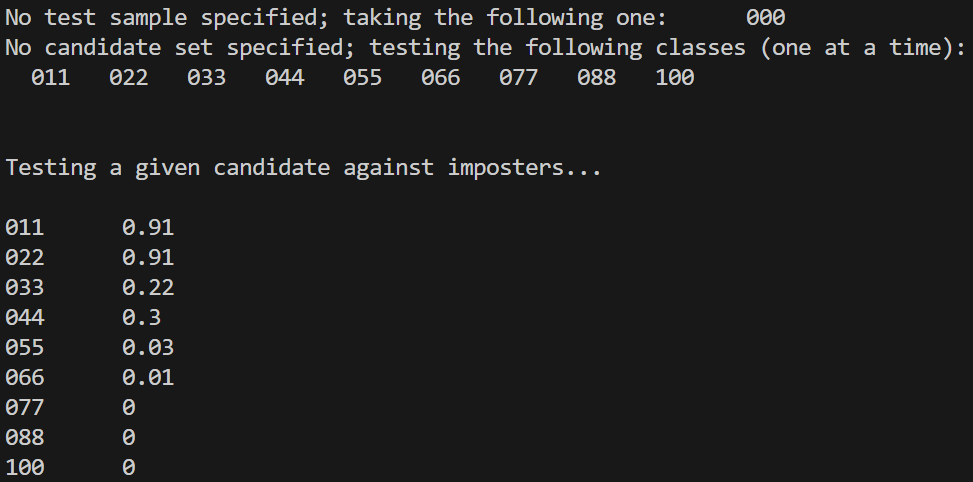}
        \caption{The \texttt{imposters()}'s results for our injection-optimality experiment.}
        \label{fig:optimality_experiment_results}
    \end{figure}

\begin{figure}[H]
        \centering
        \includegraphics[width=1\linewidth]{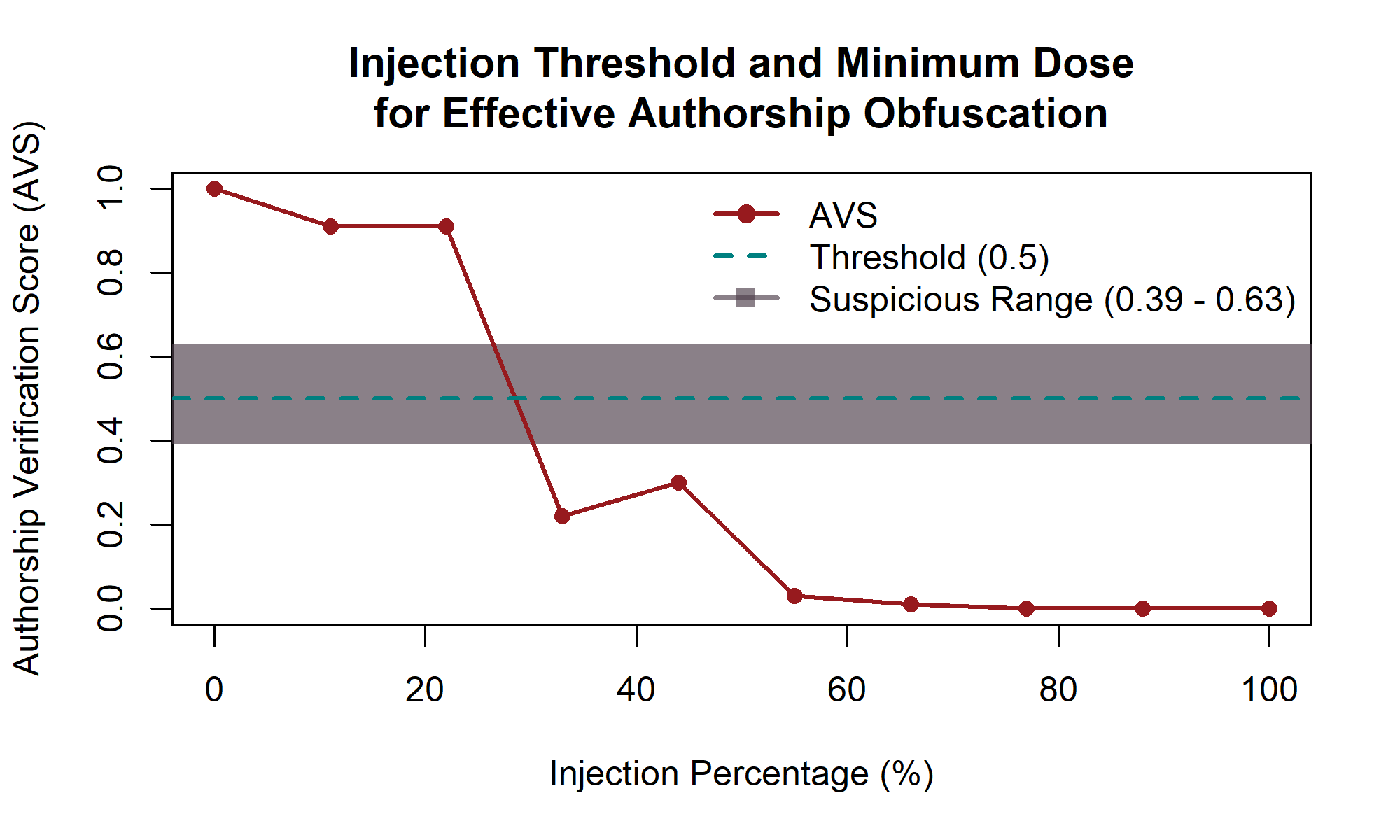}
        \caption{The relationship between injection percentage (x-axis) and authorship verification score (y-axis) as measured by \texttt{imposters()}. The curve shows the injection threshold at which the verification score drops precipitously, indicating the minimum dose required for effective authorship obfuscation.}
        \label{fig:avs_plot}
    \end{figure}

    When running these ten samples through the same authorship-verification workflow as before, we notice a trend. See (\textbf{Figure \ref{fig:optimality_experiment_results}}) for the output of the \texttt{ imposters()} results. The main takeaways are as follows. Injecting steganographic payloads with less than 22\% word coverage does little to obfuscate authorship. Based on our interpretation criterion, at this range, an author's stylistic signal remains largely intact. Anything above 33\%, however, encroaches into favorable territory; that is, 33\% coverage and above effectively ensures authorship obfuscation. Granted, our original question was one of optimality. On that front, the verdict seems to be that any coverage above 77\% yields diminishing returns, as the score remains at 0---the lowest possible value---from 77\% upward. See (\textbf{Figure \ref{fig:avs_plot}}) for a plot capturing our findings. See (\textbf{Table \ref{tab:avs_table}}) for the same information conveyed in a list format.

    \begin{table}[H]
        \centering
        \scalebox{1}{
            \setlength{\tabcolsep}{10pt}
            \begin{tabular}{|c|c|}
                \hline
        
                \rowcolor{black}
                \textcolor{white}{\textbf{Injection Percentage}} &
                \textcolor{white}{\textbf{Authorship Verification Score}} \\ 
                
                \hline

                \rowcolor{color_a} 
                \textcolor{white}{0\%} & 
                \textcolor{white}{1.00} \\
                    
                \hline

                \rowcolor{color_b} 
                \textcolor{white}{11\%} & 
                \textcolor{white}{0.91} \\
                    
                \hline

                \rowcolor{color_c} 
                \textcolor{white}{22\%} & 
                \textcolor{white}{0.91} \\
                    
                \hline

                \rowcolor{color_d} 
                \textbf{\textcolor{white}{33\%}} & 
                \textbf{\textcolor{white}{0.22}} \\
                    
                \hline

                \rowcolor{color_a} 
                \textcolor{white}{44\%} & 
                \textcolor{white}{0.30} \\
                    
                \hline

                \rowcolor{color_b} 
                \textcolor{white}{55\%} & 
                \textcolor{white}{0.03} \\
                    
                \hline

                \rowcolor{color_c} 
                \textcolor{white}{66\%} & 
                \textcolor{white}{0.01} \\
                    
                \hline

                \rowcolor{color_d} 
                \textbf{\textcolor{white}{77\%}} & 
                \textbf{\textcolor{white}{0.00}} \\
                    
                \hline

                \rowcolor{color_a} 
                \textcolor{white}{88\%} & 
                \textcolor{white}{0.00} \\
                    
                \hline

                \rowcolor{color_b} 
                \textcolor{white}{100\%} & 
                \textcolor{white}{0.00} \\
                    
                \hline

            \end{tabular}
        }
        \vspace{0.5cm}
        \caption{A tabular representation of the results from our injection-optimality experiment using \texttt{imposters()}. 33\% coverage suggests authorship obfuscation, while coverage above 77\% indicates reduced benefits.}
        \label{tab:avs_table}
    \end{table}

\section{The Privacy Implications of Stylometry}
\label{sec:Privacy_Implications}

    \epigraph{\textcolor{adversarial}{I am I, and wish I wasn't.}}{\textit{Brave New World \\ Aldous Huxley}}

    The expression \say{information wants to be free} captures the tension at the heart of stylometric analysis. While the phrase suggests that data will inevitably spread, the very act of extracting stylistic fingerprints can create an \textit{infohazard}\footnote{Nick Bostrom (\textit{Bostrom} \cite{Bostrom2010}) formally defines an information hazard (infohazard) as \say{a risk that arises from the dissemination---or the potential dissemination---of (true) information that may cause harm or enable some agent to cause harm.} Also relevant to our discussion is the idea of an \textit{enemy hazard}, which he describes as: \say{by obtaining information, our enemy---or potential---enemy becomes stronger, and this increases the threat [they pose] to us.}}: the risk that the resulting insights are weaponized against individuals. Depending on a user's motives, the patterns uncovered by stylometry may expose a writer's identity, revealing sensitive personal details such as the true source of anonymous whistleblower posts or the operational-security weaknesses of a group. This exposure can lead to legal or professional repercussions, endangering the individual's life by exposing them to prosecution, retaliation, or violent threats, and enabling the prosecution of authors of controversial or illegal content.

    Because the information meets the classic criteria for an infohazard---easy misuse and difficulty of mitigation---its disclosure carries a real danger. A leaked dataset that pairs de-anonymized whistleblower texts with the full writing samples of known individuals would allow stylometric algorithms to reliably re-identify the whistleblowers, exposing their whereabouts, connections, and motivations. Once such a dataset is public, the privacy breach cannot be easily undone.

    Assuming the most charitable circumstances, a writer's identity cannot be easily inferred from a handful of texts without a deliberate, concerted effort. The text itself does not inherently contain identifying information---or does it? By lowering the barrier to acquiring the specialized skills needed for forensic de-anonymization, stylometry transforms a previously labor-intensive task into an automated, streamlined, and widely accessible tool. This democratization amplifies both the utility of authorship verification and the potential for misuse.

    The expanded range of detectable stylistic cues means that, more often than not, a machine can detect the minutiae that pose the most concerning threat; however, a human, at face value, can only infer so much about a text's originator. The machine excels at connecting the dots, correlating evidence, and amplifying signals. Here, it is important to recognize that the machine is but a tool; it has no agency, no volition---a uniquely human trait. Blame cannot be placed squarely on the machine alone. The puppeteer ultimately decides how to manipulate the puppet, which the machine represents. Strings are pulled, and the doll is controlled to actualize the will of the puppet master. Raging against the machine may feel appropriate---viscerally so---but the machine is only complicit in the loss of freedom and rights violations that accompany stylometry.

    Without robust anonymity safeguards, a whistleblowing system, as previously illustrated, implodes in the absence of anonymity. When that protective veil disappears, grassroots activism and the speech needed for impactful communication undergo a chilling effect: a deterrent influence---often created by laws, policies, or threats---that causes individuals or groups to self-censor or refrain from exercising their rights (such as free speech, journalism, or academic inquiry) out of fear of legal repercussion, retaliation, or social backlash. The effect \say{cools} the willingness to speak or act, even when no actual penalty has been imposed.

    Thus, anonymity, once a formidable shield, now faces the shield-breaker of stylometry. No longer can an online epithet or superficial obfuscation technique conceal one's identity. The contents of books, articles, emails, texts, and social-media posts---and the information that can be stylometrically extracted from them---now constitute a monumental vulnerability, expanding the existing attack surface and facilitating further privacy invasions.
    
    While self-immolation remains viable\footnote{Disclaimer: We do not encourage, endorse, or recommend any illegal, harmful, or self-destructive actions. Readers are solely responsible for their own actions and any consequences that may arise.}---a tactic some might consider to evade forensic identification or prevent tracking by scorching away their \say{physical fingerprint}---the \say{cognitive fingerprint} that stylometry targets is not made of similarly combustible stuff. To put it another way, physical attributes can be burned away, erased, or concealed with relative ease, but the patterns embedded in a person's language, syntax, word choice, rhythm, and underlying thought processes are far more resilient. These linguistic signatures persist across topics, platforms, and even after a writer attempts to disguise their style, because they stem from deeply ingrained habits, education, cultural background, and neurological wiring. While one can destroy the visible marks on a body, the invisible marks left by the mind remain far harder to extinguish. As such, proactive measures must be taken to suppress this cognitive fingerprint, the distinctive pattern of mental and linguistic traits a writer consistently exhibits. This undertaking, for all intents and purposes, is tantamount to self-suppression.

    Like a certain self-inflicted, pyro-induced wound made in protest, self-suppres\-sion may seem an acceptable sacrifice, but the stakes rise dramatically when profiling extends beyond authorship. Formulating a profile to prove whether an author wrote a book pales in comparison to the potential of identifying a health condition, such as a mental disorder, from online writings. Detecting a writer's preferences, views, and ideologies is also within stylometry's purview, as is linking profiles across different social networks. While stylometry offers many benefits, it would be disingenuous to ignore the egregious human-rights violations it can enable. Personal data---text data, specifically---is intimate and wholly deserving of privacy, regardless of its source. In this regard, \textit{Patergianakis et al.} \cite{Patergianakis2022} make a poignant point: \say{If a user assumes that they remain anonymous in a specific context, but a stylometric technique suffices, by reasonable means, to identify them or to link them with another electronic account, then their data becomes erroneously considered anonymous [or pseudonymous]\dots [and r]e-identification may allow further processing of their data for [privacy-violating] purposes.}

    With this in mind, stoking fear of stylometry, in our opinion, is not fearmongering. We are deliberately amplifying concerns that an individual's writing style can be identified, tracked, or profiled by computational analysis. We wish to tap into anxieties about privacy, surveillance, and the loss of anonymity. Every word leaves a traceable \say{digital fingerprint} that could be used to expose identity, intentions, or affiliations.

    Under these circumstances, individuals and marginalized groups often lack the resources needed to protect their data, whereas corporations, governments, and other powerful actors have the ability to shape the rules, keeping their own deeds hidden \cite{DilworthGwen2026}.

    Adversarial stylometry is a means to take back the reins: \textit{privacy for the weak, transparency for the powerful.}
    
\bibliographystyle{splncs04}
\bibliography{StegoStylo.bib}

\begin{thebibliography}{10}
\providecommand{\url}[1]{\texttt{#1}}
\providecommand{\urlprefix}{URL }
\providecommand{\doi}[1]{https://doi.org/#1}

\bibitem{Imposters2025}
{Authorship Verification Classifier Known as the Imposters Method}. R
  Documentation
  \url{https://search.r-project.org/CRAN/refmans/stylo/html/imposters.html}

\bibitem{LinuxBash2025}
{Match invisible Unicode characters (eg, zero-width spaces) using `grep -P'}.
  Linux Bash  (3 2025),
  \url{https://www.linuxbash.sh/post/match-invisible-unicode-characters-eg-zero-width-spaces-using-grep--p}

\bibitem{Abrosimov2025}
Abrosimov, K., Grebennikov, A., Tzanetakis, G., Sidorova, A.: {Linguistic Tools
  in Musical Stylometry}. Anthology of Computers and the Humanities
  \textbf{3},  641--652 (11 2025). \doi{10.63744/Av1c2rVmcj0N},
  \url{https://anthology.ach.org/volumes/vol0003/linguistic-tools-in-musical-stylometry/}

\bibitem{Albaqami2026}
Albaqami, H., Ayub, M.A., Ahmad, N., Ahmad, Y., Alqahtani, M.M., Algamdi, A.M.,
  Owaidah, A.A., Ahmad, K.: {Stylometry Analysis of Human and Machine Text for
  Academic Integrity}  (1 2026), \url{https://arxiv.org/abs/2601.01225}

\bibitem{Meerza2025}
Ali~Meerza, S.I., Sun, L., Liu, J.: {Harmonycloak: Making Music Unlearnable for
  Generative AI}. In: 2025 IEEE Symposium on Security and Privacy (SP). pp.
  430--448 (2025). \doi{10.1109/SP61157.2025.00085},
  \url{https://ieeexplore.ieee.org/document/11023354}

\bibitem{Barrios2023}
Barrios, J., Gabay, S., Cafiero, F., Debban\'e, M.: {Detecting Psychological
  Disorders with Stylometry: the Case of ADHD in Adolescent Autobiographical
  Narratives}  (10 2023). \doi{10.31234/osf.io/s5cm3},
  \url{https://osf.io/preprints/psyarxiv/s5cm3_v1}

\bibitem{Bostrom2010}
Bostrom, N.: {Information Hazards: A Typology of Potential Harms from
  Knowledge}. Review of Contemporary Philosophy  \textbf{10},  44--79 (2010),
  \url{https://nickbostrom.com/information-hazards.pdf}

\bibitem{DilworthGwen2026}
Dilworth, G.: {Constituents are seldom heard in the Mississippi Legislature.
  Legal experts say easy fixes could amplify people's voices}. Mississippi
  Today  (1 2026),
  \url{https://mississippitoday.org/2026/01/01/constituents-mississippi-legislature/}

\bibitem{TraceTarnish2025}
Dilworth, R.: {Tuning for TraceTarnish: Techniques, Trends, and Testing
  Tangible Traits}  (12 2025), \url{https://arxiv.org/abs/2512.03465}

\bibitem{AdversarialStylometry2025}
Dilworth, R.: {Unveiling Unicode's Unseen Underpinnings in Undermining
  Authorship Attribution}  (10 2025), \url{https://arxiv.org/abs/2508.15840}

\bibitem{Eder2018}
Eder, M.: {Authorship verification with the package `stylo'}. Computational
  Stylistics Group  (5 2018),
  \url{https://computationalstylistics.github.io/blog/imposters/}

\bibitem{translateLocally2021}
Heafield, K., Bogoychev, N., Nail, G., van~der Linde, J.: {translateLocally}.
  GitHub  (2021), \url{https://github.com/XapaJIaMnu/translateLocally},
  \url{https://private.mt/}

\bibitem{Jordan2025}
Jordan, B.: {The Art Of Poison-Pilling Music Files}. YouTube  (4 2025),
  \url{https://www.youtube.com/watch?v=xMYm2d9bmEA}

\bibitem{Koppel2017}
Koppel, M., Seidman, S.: {Detecting pseudepigraphic texts using novel
  similarity measures}. Digital Scholarship in the Humanities  \textbf{33},
  72--81 (12 2017). \doi{10.1093/llc/fqx011},
  \url{https://doi.org/10.1093/llc/fqx011}

\bibitem{Lerner2025}
Lerner, R.: {How to Cyberstalk -- A Conversation about Adversarial Stylometry}
  (7 2025),
  \url{https://robert-lerner.com/how-to-cyberstalk-a-conversation-about-adversarial-stylometry/}

\bibitem{Mikros2025}
Mikros, G.: {Large Language Models and Forensic Linguistics: Navigating
  Opportunities and Threats in the Age of Generative AI}  (12 2025),
  \url{https://arxiv.org/abs/2512.06922}

\bibitem{Opara2025}
Opara, C., Modesti, P., Golightly, L.: {Evaluating spam filters and Stylometric
  Detection of AI-generated phishing emails}. Expert Systems with Applications
  \textbf{276},  127044 (6 2025). \doi{10.1016/j.eswa.2025.127044},
  \url{https://www.sciencedirect.com/science/article/pii/S0957417425006669}

\bibitem{Patergianakis2022}
Patergianakis, A., Limniotis, K.: {Privacy Issues in Stylometric Methods}.
  Cryptography  \textbf{6}, ~17 (4 2022). \doi{10.3390/cryptography6020017},
  \url{https://www.mdpi.com/2410-387X/6/2/17}

\bibitem{Perez2022}
P\'erez, A., Parapar, J., Barreiro, A.: {Automatic depression score estimation
  with word embedding models}. Artificial Intelligence in Medicine
  \textbf{132},  102380 (10 2022). \doi{10.1016/j.artmed.2022.102380},
  \url{https://www.sciencedirect.com/science/article/pii/S093336572200135X}

\bibitem{Rezaii2019}
Rezaii, N., Walker, E., Wolff, P.: {A machine learning approach to predicting
  psychosis using semantic density and latent content analysis}. npj
  Schizophrenia  \textbf{5}, ~9 (6 2019). \doi{10.1038/s41537-019-0077-9},
  \url{https://www.nature.com/articles/s41537-019-0077-9}

\bibitem{LLMask2024}
top{-}on: {LLMask}. GitHub  (2024), \url{https://github.com/top-on/llmask}

\bibitem{Yang2024}
Yang, H.C., Hung, Y.L., Wang, L.C.: {Stylometry-based Fake News Classification
  Using Text Mining Techniques}. In: Proceedings of the 2024 11th
  Multidisciplinary International Social Networks Conference. pp. 85--94. ACM
  (8 2024). \doi{10.1145/3675669.3675682},
  \url{https://dl.acm.org/doi/10.1145/3675669.3675682}

\end{thebibliography}

\end{document}